\documentclass[manuscript,screen]{acmart}

\AtBeginDocument{%
  }
    
\renewcommand\footnotetextcopyrightpermission[1]{}    

\usepackage{tcolorbox}
\usepackage{xspace}
\usepackage{flushend}
\usepackage{enumitem}
\usepackage{algorithm}
\usepackage{listings}
\usepackage{xcolor}
\usepackage{algorithm}
\usepackage{amsthm}
\theoremstyle{acmdefinition}

\usepackage{multirow}
\usepackage{booktabs}
\usepackage{tabularx}
\newcolumntype{C}[1]{>{\centering\arraybackslash}p{#1}}

\usepackage{subcaption}

\usepackage{nicefrac, xfrac}

\usepackage{microtype}
\usepackage{tikz}
\definecolor{s1}{RGB}{179,218,219}
\definecolor{s2}{RGB}{82,122,154}
\definecolor{s3}{RGB}{36,53,84}

\newcommand{\todo}[1]{\textcolor{red}{#1}}

\newcommand{\ie}[0]{\textit{i.e.,}\xspace}
\newcommand{\eg}[0]{\textit{e.g.,}\xspace}

\newcommand{\huangyiheng}[1]{\textcolor{black}{#1}} 
\definecolor{zztcolor}{rgb}{0.6,0.1,0.8}

\definecolor{darkgreen}{rgb}{0.0, 0.4, 0.0}
\newcommand{\wss}[1]{\textcolor{black}{#1}} 

\begin{document}

\title{Lifting the Veil on Composition, Risks, and Mitigations of the Large Language Model Supply Chain}


\author{Kaifeng Huang}

\affiliation{%
  \institution{Tongji University}
  \city{Shanghai}
  \country{China}
}

\author{Bihuan Chen}
\affiliation{%
  \institution{Fudan University}
  \city{Shanghai}
  \country{China}
}

\author{You Lu}
\affiliation{%
\authornotemark[2]
  \institution{Fudan University}
  \city{Shanghai}
  \country{China}
}

\author{Susheng Wu}
\affiliation{%
  \institution{Fudan University}
  \city{Shanghai}
  \country{China}
}

\author{Dingji Wang}
\affiliation{%
  \institution{Fudan University}
  \city{Shanghai}
  \country{China}
}

\author{Yiheng Huang}
\affiliation{%
  \institution{Fudan University}
  \city{Shanghai}
  \country{China}
}

\author{Haowen Jiang}
\affiliation{%
  \institution{Fudan University}
  \city{Shanghai}
  \country{China}
}

\author{Zhuotong Zhou}
\affiliation{%
  \institution{Fudan University}
  \city{Shanghai}
  \country{China}
}

\author{Junming Cao}
\affiliation{%
  \institution{Fudan University}
  \city{Shanghai}
  \country{China}
}

\author{Xin Peng}
\affiliation{%
  \institution{Fudan University}
  \city{Shanghai}
  \country{China}
}

\renewcommand{\shortauthors}{Huang et al.}



\begin{abstract}

Large language models (LLMs) have sparked significant impact with regard to both intelligence and productivity. Numerous enterprises have integrated LLMs into their applications to solve their own domain-specific tasks. However, integrating LLMs into specific scenarios is a systematic process that involves substantial components, which are collectively referred to as the \textit{LLM supply chain}. A comprehensive understanding of LLM supply chain composition, as well as the relationships among its components, is crucial for enabling effective mitigation measures for different related risks. While existing literature has explored various risks associated with LLMs, there remains a notable gap in systematically characterizing the LLM supply chain from the dual perspectives of contributors and consumers. In this work, we develop a structured taxonomy encompassing risk types, risky actions, and corresponding mitigations across different stakeholders and components of the supply chain. We believe that a thorough review of the LLM supply chain composition, along with its inherent risks and mitigation measures, would be valuable for industry practitioners to avoid potential damages and losses, and enlightening for academic researchers to rethink existing approaches and explore new avenues of research.

\end{abstract}

\maketitle


\section{Introduction}

The large language models (LLMs) have sparked unprecedented discussion about the power of generative language models. There has been a significant surge in the development and introduction of large language models, regarding both commercial LLMs and free open-source LLMs. Researchers and practitioners have extended the capabilities of LLMs beyond natural language processing, applying them in fields like software engineering, finance, and education, dramatically reshaping these areas~\cite{llm2023year, fan2023large, bloomberg}.

Integrating LLMs involves more than just utilizing the models themselves. On the one hand, although some applications may directly employ commercial LLMs, they still require additional solutions such as pre-processing (\eg Promptify \cite{promptify} for prompt engineering, GPTCache \cite{gptcache} for reducing LLM API call expenses), etc. 
On the other hand, businesses may prefer open-source LLMs to enable flexible customizations, which demand more sophisticated components, during stages such as fine-tuning (\eg Labelbox \cite{labelbox}), model conversion (\eg Onnx \cite{onnx}), quantization (\eg TensorRT \cite{tensorrt}), etc. 
Consequently, there exists a multitude of sub-components that are transitively depended upon. Furthermore, the composition types in the \textit{LLM supply chain} are more varied than in the traditional software supply chain. Overall, leveraging LLMs is a systematic process that entails numerous components and composition types, collectively referred to as the \textit{LLM supply chain}.


Accordingly, we are confronted with a fresh LLM supply chain ecosystem, where the attack surface has expanded but remains understudied, and the number of participants is rapidly increasing. While the LLM supply chain shares similarities with the well-recognized open-source software supply chain \cite{ladisa2023sok}, it also exhibits its uniqueness. For instance, attackers can exploit publicly accessible LLMs to extract sensitive information (\eg privacy leakage) or fine-tune open-source LLMs \cite{openllms} to embed malicious instructions (\eg backdoor attacks). 




 \begin{figure*}[!t]
    \centering
    \includegraphics[width=0.95\textwidth]{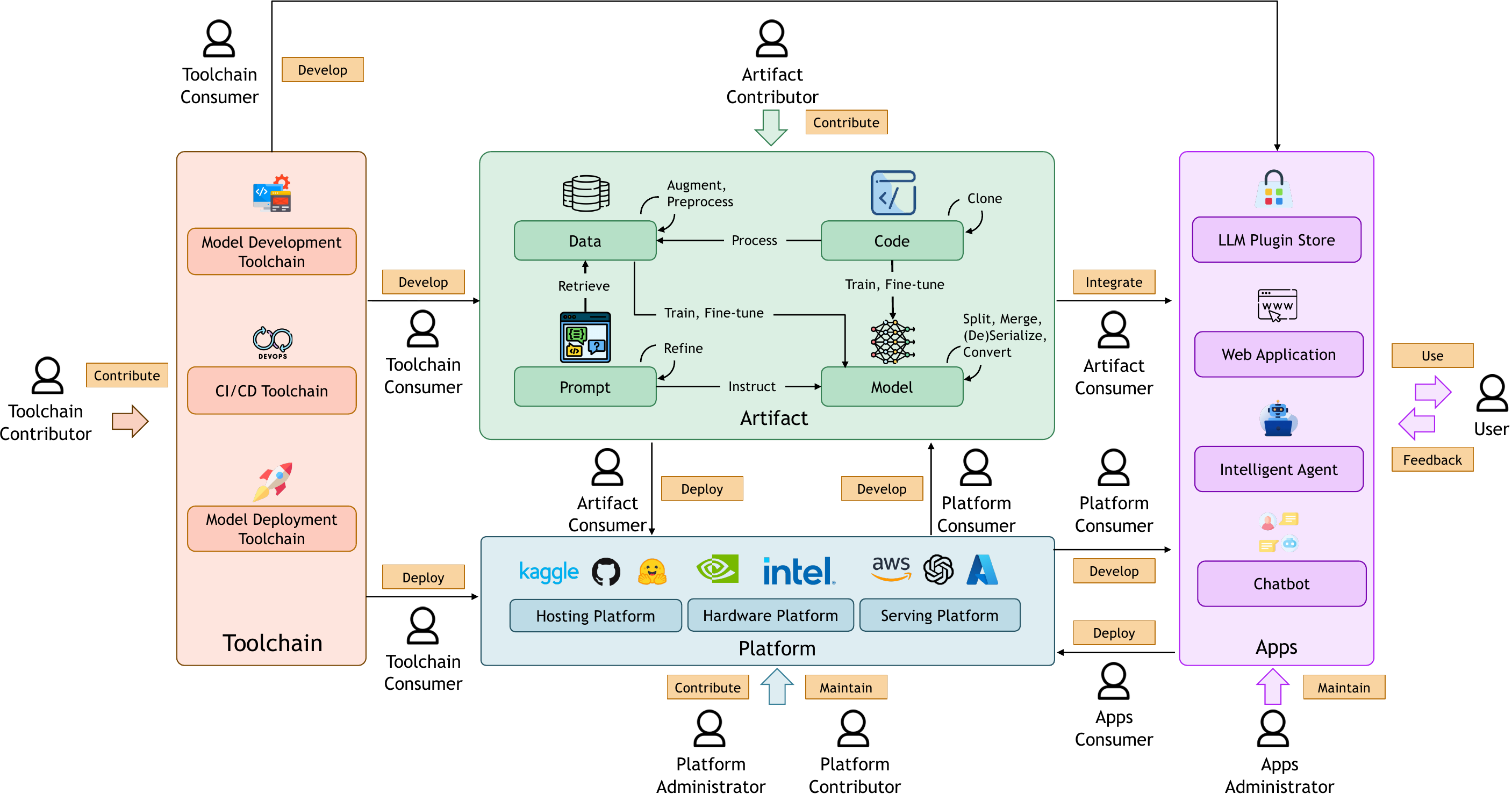}
    \caption{Overview of Large Language Model Supply Chain}
    \label{fig:overview}
  \end{figure*}

We provide our definition for the large language model supply chain, as shown in Fig. \ref{fig:overview}. It encompasses the ecosystem and sequence of processes integral to developing, training, deploying, and distributing applications that leverage large language models, including: 
\begin{itemize}[leftmargin=*]
  \item \textit{Stakeholders in various roles}. \eg developers contributing to the artifacts, organizations managing the platforms, or vendors providing cloud computing services.
  \item \textit{Upstream artifacts in multiple forms}. \eg plaintext source code, binary executables, or serialized models.
  \item \textit{Diverse supply relationships}. \eg augmentation relationships for data, cloning relationships for code, and merging relationships for models.
  \item \textit{Development stages for various purposes}. \eg data preprocessing, model pre-training, fine-tuning, integration, and monitoring.
\end{itemize}

{While existing literature discusses risks associated with LLMs, there rarely exists a comprehensive work that outlines the LLM supply chain from the perspective of contributors and consumers. } {We aim to answers the following research questions:}

\noindent {\textbf{RQ1 Composition Analysis:} What new components accompanied with their defining characteristics, emerge in the LLM supply chain, compared with the software supply chain?} 

\noindent {\textbf{RQ2 Risk Analysis:} How to characterize the risks posed by the LLM supply chain? What are their uniqueness, compared with the software supply chain?}

\noindent {\textbf{RQ3 Mitigation Analysis:} How to mitigate the risks in the LLM supply chain? }

To study \textbf{RQ1}, we provide a comprehensive overview of the LLM supply chain, detailing the stakeholders, composing artifacts, and the supplying types, as illustrated in Figure~\ref{fig:overview}. Our definition of the LLM supply chain differs from Want et al.'s work \cite{wang2024large2} in emphasizing the LLM supply chain more on the basic components and their supply relationships between them. 
To explore \textbf{RQ2}, we developed taxonomies of risk types, risky actions, and mitigations related to various LLM supply chain stakeholders and components. Specifically, we fist holistically collected both academic and online resources. Then, we summarize the risk scenarios as \textit{stakeholders} performing \textit{risky actions} on specific \textit{supply chain components}, which in turn lead to distinct \textit{risk types}. We illustrate the risk scenarios from the perspective of upstream contributors, downstream users, and administrators. Furthermore, to study \textbf{RQ3}, we list the types of various mitigation measures in response to the LLM supply chain risks.

Our paper explores the technical and operational aspects of the \textit{LLM supply chain}, offering a holistic understanding that draws valuable insights for researchers and engineers in software engineering, system architecture, software security, and data governance. The contributions of our paper are as follows: 

\begin{itemize}[leftmargin=*]
\item We presented a comprehensive overview of the \textit{LLM supply chain}, including artifacts, stakeholders, and composition types. 
\item We develop a detailed taxonomy of risk and mitigations with regard to stakeholders, risk types and supply chain artifacts, providing actionable guidance for practitioners involved in the LLM supply chain.
\item We envision future challenges and opportunities in securing the LLM supply chain.
\end{itemize}



\section{Related Work}\label{related_work}

Various security-related aspects of large language models are discussed in existing work. For ease of comparison, we focus our topics in \textit{attacks and risks} and \textit{defensive techniques}. Additionally, we unified the terminology and definitions across the surveys to enable a clearer comparison.

\textbf{Attack and Risks.} In the context of attacks and risks, the \textit{Output Risk} is extensively covered in eight survey papers, which refers to the potential for LLMs to generate harmful, untruthful, hallucinatory, or unhelpful content. Following this, \textit{Privacy Attacks} are another major concern, with nine papers addressing various aspects such as membership inference attacks, privacy leakage (e.g., PPI) attacks, gradient leakage attacks, model inversion attacks, and attribute inference attacks. Besides, \textit{Prompts Attack} is covered in seven papers, including prompt injection attack and jailbreaking attack. 
Unfortunately, only a few papers \cite{pankajakshan2024mapping, cui2024risk} mention \textit{Toolchain Attacks}. Yao et al. \cite{yao2024survey} discuss \textit{Supply Chain Vulnerabilities}, and Wu et al. \cite{wu2024new} highlight issues like malicious webtool misuse and cross-session access, which are closely related but account for a subset of \textit{Toolchain Attacks}. Therefore, the topic of toolchain attacks was not explored in depth. Finally, only Neel et al. \cite{neel2023privacy} addressed the issue of \textit{Copyright Risks}.

\textbf{Defenses and Mitigation.} Defenses and Mitigation are suggested in accordance with the corresponding attacks and risks. For \textit{Prompt Attacks}, existing works propose \textit{Input Sanitization} \cite{pankajakshan2024mapping, iqbal2023llm}. To mitigate \textit{Output Risk}, approaches like \textit{Output Sanitization} are recommended by several studies \cite{yao2024survey, cui2024risk, kumar2024ethics}. For \textit{Data Attacks}, the literature suggests techniques such as \textit{Data Cleaning} \cite{yao2024survey}, \textit{Data De-duplication} \cite{neel2023privacy}, and \textit{Data Sanitization} \cite{esmradi2023comprehensive}. However, these defensive techniques are not well-summarized across the surveyed works. While some methods, such as \textit{Differential Privacy}, \textit{Federated Learning}, or \textit{LLM Alignment}, refer to specific technical measures designed to defend against particular attacks or mitigate associated risks, other defensive measures are less detailed. For instance, strategies like \textit{Output Detection} and \textit{Input Sanitization} are suggested to address \textit{Output Risk} and \textit{Prompt Attacks}, respectively, but they lack in-depth technical details on how these defenses are implemented.

\textbf{Comparison to Existing Surveys.} Our paper contributes the following contributions which are novel and incremental to existing surveys. First, we conducted a comprehensive survey in both academic literature and online resources regarding the \textit{open-source LLM supply chain}. Second, we provide a comprehensive overview and highlight the key stakeholders and components in the LLM supply chain. Our definition of the LLM supply chain differs from Want et al.'s work \cite{wang2024large2} in emphasizing the LLM supply chain more on the basic components and their supply relationships between them.  Third, we summarize the risk scenarios as \textit{stakeholders} performing \textit{risky actions} on specific \textit{supply chain components}, as well as \textit{risk types}. We illustrate the risk scenarios from the perspective of three major stakeholders. Furthermore, we list the types of various mitigation measures in response to the LLM supply chain risks.


\section{Methodology}\label{sec:approach}

\begin{figure}[!t]
  \centering
  \includegraphics[width=0.60\textwidth]{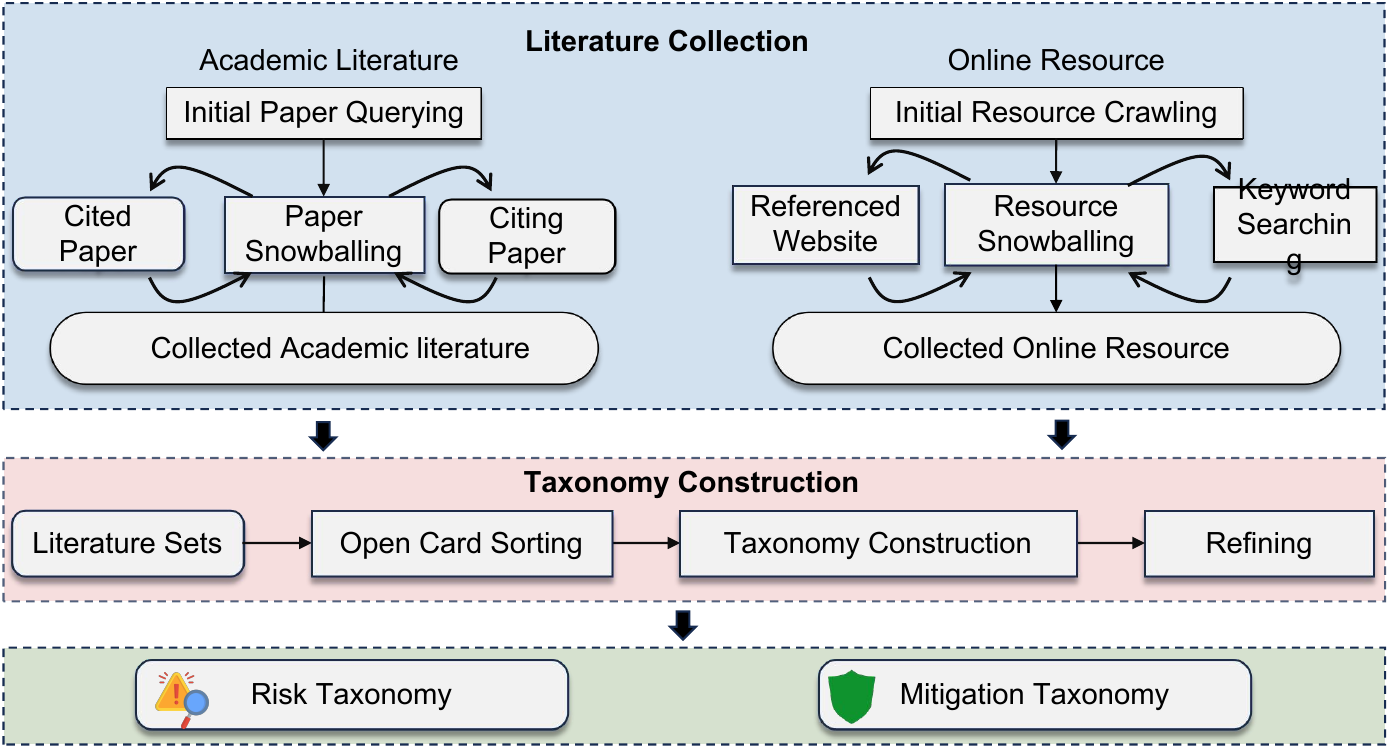}
  \caption{Approach Overview of our Literature Review}
  \label{fig:approach}
\end{figure}

Following the systematic literature review guidelines~\cite{garousi2016systematic}, we adopt a scientific and systematic approach for literature collection (see Sec.~\ref{sec:approach:collect}), then we performed a taxonomy analysis (see Sec.~\ref{sec:TaxonomyConstruction}). An overview of our review process is presented in Figure~\ref{fig:approach}.

\subsection{Literature Collection}\label{sec:approach:collect}
We review the academic literature and online resources to collect an extensive list of risks and mitigations about LLM supply chain.

\subsubsection{Academic Literature Collection}
We define the temporal scope of our survey to span from the emergence of the earliest relevant publications in this field to the most recent contributions, specifically covering the period from January 2018~to~December 2024 (\ie around 7 years). Two PhD students with over five years of experience in software supply chain management and two PhD students with over three years in LLM security have engaged in this process. The detailed steps are as follows.

\textbf{Initial Paper Querying.} We identify potentially relevant papers by conducting searches across major electronic scholarly databases. Specifically, we select ACM Digital Library\footnote{https://dl.acm.org/}, IEEE Xplore\footnote{https://ieeexplore.ieee.org/Xplore/home.jsp}, DBLP\footnote{https://dblp.org/} and Semantic Scholar\footnote{https://www.semanticscholar.org/me/research} as our primary data sources, which are popular bibliography databases containing a comprehensive list of research venues in computer science. Then, we define the keywords presented in Listing~\ref{listing} from three domains, \ie supply chain security (\eg security, attack, malware, threat, vulnerabilities), LLM security (\eg data poisoning, data provenance, model poisoning) and LLM DevOps (\eg data validation, model validation, model evaluation).  

\lstset{
  basicstyle=\ttfamily,
  breaklines=true,
  backgroundcolor=\color{white},
  frame=single,
  keywordstyle=\color{blue},
  commentstyle=\color{blue},
  stringstyle=\color{blue},
}

\begin{lstlisting}[language=SQL, caption={Search Keywords for Literature Collection},label=listing]
  (LLM security AND supply chain security)
  OR (LLM security AND LLM DevOps) OR
  (supply chain security AND LLM DevOps)
  OR LLM security
\end{lstlisting}

Furthermore, we perform a manual assessment to determine the relevance of each candidate paper to the topic of LLM supply chain, excluding those that do not align with the goal of our study. Besides, we exclude short papers with fewer than 2 pages, as well as survey paper papers or summary papers, which are separately discussed in Sec.~\ref{related_work} for a comparison with our work. Consequently, we obtain an initial set of \todo{127} papers.

\textbf{Paper Snowballing.} To mitigate the risk of omitting relevant papers, we apply both backward and forward snowballing techniques~\cite{wohlin2014guidelines} on the initial paper set. In the backward snowballing phase, we check the reference lists of these papers to discover new candidates. In forward snowballing phase, we use Google Scholar\footnote{https://scholar.google.com/} to identify papers that cite the initial set. This iterative process is repeated until no further relevant papers are found. All newly identified candidate papers are also manually assessed for relevance, resulting in a final set of \todo{238} papers.

\subsubsection{\wss{Online Resource Collection}} \huangyiheng{In addition to collecting papers in the academic literature, we also include online resources, \eg blog posts and incident reports, which often include real-world attacks or the latest knowledge not yet covered by academic literature.} 
These websites are selected based on their broad coverage, strong reputation in the cybersecurity community, professional content, timely updates, and authoritative reporting. Following a procedure similar to that used for academic literature collection, on the one hand, we leverage the same keywords used in the academic literature collection on the websites containing querying interfaces. On the other hand, we manually inspect the websites to collect the relevant news or events, resulting in \todo{31} reports related to our topic.


\subsection{Taxonomy Construction}\label{sec:TaxonomyConstruction}
We download all the relevant resources and conduct a full-text analysis to inspect and categorize various elements of large language model risks and mitigations from the literature. First of all, we use open card sorting in our analysis to highlight terms, concepts, techniques, and approaches related to the large language model risks and mitigations. Then, we perform the hierarchical tree structure construction, where we connect the identified elements into a structured, hierarchical framework that reflects the relationships between different aspects of the LLM supply chain. We organize tree structures where the root represents the risks, stakeholders, and mitigations. At last, we refine our summarized structures according to the following criteria with multiple iterations.

\begin{itemize}[leftmargin=*]
    \item \textbf{Hierarchy.} Evaluate whether the tree's structure is logical, ensuring broader attack categories at higher levels and specific methods at lower levels.
    \item \textbf{Accuracy.} Verify the precision of node descriptions by consulting attack case studies and expert feedback, ensuring alignment with real-world attack patterns.
    \item \textbf{Comprehensibility.} Test the clarity of node descriptions through expert panels or user feedback, ensuring that they are easy to understand.
    \item \textbf{Relevance.} Ensure that each node reflects the latest risks and mitigations by reviewing recent research and industry standards. 
    \item \textbf{Mutual Exclusivity.} Assess whether each node is distinct from the others. 
    \item \textbf{Coverage.} Check whether the node includes all known attack techniques for its type, comparing against existing security frameworks and databases.
\end{itemize}

Each node in the structures is assessed by the four professionals independently. The structures are refined and re-evaluated iteratively until all professionals reach a final agreement. Furthermore, the label is left blank if the literature does not cover specific aspects (\eg mitigation).

\section{Composition of Large Language Model Supply Chain (RQ1)}\label{sec:component}

We present an overview of the large language model supply chain in Figure \ref{fig:overview}. 
In general, we identify four types of \textit{stakeholders}, interacting with specific \textit{components} of the large language model supply chain.

\subsection{Stakeholders}\label{sec:stakeholder}

We summarize four roles of stakeholders in the LLM supply chain. i.e., \textit{contributors}, \textit{consumers}, \textit{administrators}, and \textit{user}. 

\textbf{Contributors.} Person who engaged in development activities and sharing artifacts, such as data, models, prompts, or third-party libraries, etc. \textit{Platform Contributors} are responsible for contributing to hosting platforms, hardware platforms, and serving platforms (\ie Amazon Web Services \cite{aws_ami}) and providing stable, isolated, and managed environments. 
\textit{Artifact Contributors} process, and upload various forms of data, code, prompts, and models to the hosting platforms. They can be data engineers, model developers, prompt engineers, and library developers. \textit{Toolchain Contributors} are responsible for developing, maintaining, and improving the tools used throughout the model lifecycle. \eg Jupyter, TensorFlow for model development, Jenkins, GitHub Actions in CI/CD.

\textbf{Consumers.} We distinguish consumers depending on their consuming components, including platform consumers, artifact consumers, toolchain consumers, and app consumers. \textit{Platform Consumers} are entities that use the platform infrastructure for various purposes, such as accessing hosting services, utilizing computational resources, or interacting with serving platforms for deploying and running models and applications. \textit{Artifact Consumers} are those who utilize the artifacts produced and shared within the platform. This includes data sets, code, prompts, and models. \textit{Toolchain Consumers} are individuals or teams that leverage the toolchain components for developing, deploying, and maintaining models. They use model development, CI/CD, and deployment tools to streamline the lifecycle of models and applications. \textit{App Consumers} are end-users or other systems that interact with applications built using the platform's resources and artifacts. 

\begin{figure}[!t]
  \centering
  \includegraphics[width=0.75\textwidth]{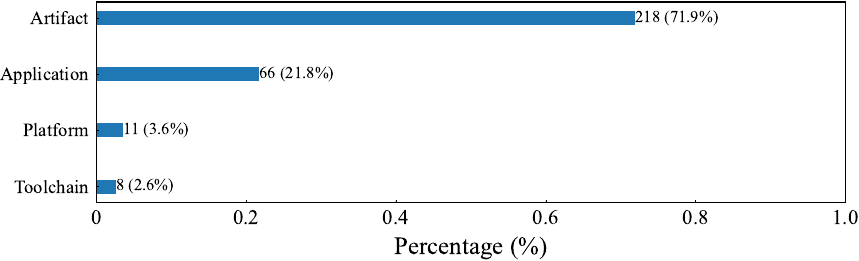}
  \caption{Distribution of Components that Covered in our Collected Literature}
  \label{fig:compoisitionbar}
\end{figure}

\textbf{Users.} The users of LLM applications perform interactions with LLM applications in various forms. Their feedbacks (explicitly or implicitly) may be collected back to the LLM application.

\textbf{Administrators.} We summarize administrators as platform administrators and apps administrators.
\textit{Platform Administrators} represent a group of people who manage and oversee the infrastructure and another group who manage repositories of artifacts. They hold the privileges to merge pull requests, manage versioning, and onboard new contributors.
\textit{Apps Administrator} is responsible for managing the lifecycle of applications that rely on the artifacts and models within the platform. This includes deploying, configuring, and monitoring applications.

\subsection{Components Analysis}

We summarize four components, including \textit{artifact}, \textit{toolchain}, \textit{platform}, and \textit{LLM~applications} (\textit{apps}). Figure \ref{fig:compoisitionbar} illustrates the distribution of supply chain components covered in the surveyed literature. Notably, the majority of the works (218 papers, 71.9\%) focus on risks associated with \textit{artifacts}, followed by \textit{applications} (66 papers, 21.8\%). In contrast, significantly fewer studies address risks related to the \textit{platform} (11 papers, 3.6\%) and \textit{toolchain} (8 papers, 2.6\%). Given that the \textit{platform} and \textit{toolchain} serve as foundational and widely adopted components in the LLM supply chain, their limited coverage in existing research underscores a critical gap and highlights the need for greater research attention.

\subsubsection{Artifacts}\label{sec:artifacts}

The LLM supply chain artifacts include data, model, prompts and code.

\textbf{Data.} The data can be divided as the corpus for pre-training or fine-tuning, and the knowledge database for retrieval augmented generation.

\textbf{Model.} The model represents the large language model that is trained and fine-tuned. The model may be adjusted or modified by splitting, merging, serializing, or deserializing components as needed to optimize performance. 

\textbf{Prompt.} The prompt is crafted to instruct the model effectively, guiding it on what information to retrieve, process, or generate. 

\textbf{Code.} The code encompasses the software and scripts utilized for data processing, model training, and prompt refinement. This includes code that directly invokes relevant library APIs, as well as the libraries themselves as dependencies. 

\subsubsection{Platforms}\label{sec:platforms}

The platforms are publicly accessible websites that provide essential infrastructure and services for distributing, and deploying various components of the LLM~ecosystem.

\textbf{Hosting Platform}. These platforms facilitate the storage, and sharing of software packages, models, data, and related resources required for LLM development and deployment. \eg Data Hosting Platforms (\eg Kaggle \cite{kaggle}), Model Hosting Platforms (\eg Hugging Face \cite{hugging_face}), Repository Hosting Platforms (\eg GitHub \cite{github}), Package Registries (\eg PyPI \cite{pypi}), Plugin Marketplaces (e.g., Visual Studio Marketplace \cite{vscode_marketplace}).

\textbf{Hardware Platform}. These platforms consist of specialized hardware designed to optimize the performance of LLMs. 

\textbf{Serving Platform}. These platforms offer cloud-based services to support the training, deployment, and maintenance of LLMs. 

\subsubsection{Toolchain}\label{sec:toolchain}
The development and deployment of large language models (LLMs) involve two connected toolchains: model development toolchain and model deployment toolchain.

\textbf{Model Development Toolchain.} Development toolchain encompasses the tools and workflows used to train, refine, and optimize the model. Key components include training frameworks, model serialization and versioning, model conversion and model quantization tools.

\textbf{Model Deployment Toolchain.} Deployment toolchain focuses on integrating the trained model into production environments throughout the model's lifecycle in production. It automates deployment, monitoring, and real-time updates. Core components include CI/CD systems, which ensure seamless integration and continuous updates in live environments, and cloud services or containerization tools (e.g., Docker) that provide the deployment environment.

\subsubsection{Applications}\label{sec:apps}

Applications provide functionalities and services directly used by users. These applications include chatbots, web applications, intelligent agents, text-generation tools, and more. Application developers integrate large models into real-world products to develop intelligent applications with API interfaces. For example, the LLM plugin store can offer a wide range of customizable plugins that allow developers to easily access and implement various functionalities tailored to specific needs. This ecosystem is examined in detail in Zhao et al.'s forward-looking analysis of the LLM app store \cite{zhao2024llm}.
 
\begin{tcolorbox}[size=title, opacityfill=0.15]
 \textbf{Summary.} We examined four key components in the LLM supply chain: artifacts, platforms, toolchains,~and applications. These components can be further broken down~into four types of artifacts, three categories of platforms, three toolchain types, and four applications. While most studies focus on risks related to \textit{artifacts} and \textit{applications}, the limited attention to the widely adopted and foundational components, i.e., \textit{platform} and \textit{toolchain}, reveals a critical research gap in the LLM supply chain.
 \end{tcolorbox}

\subsection{Composition Relation Discussion}

We discuss the composition of relations within the LLM supply chain from the perspectives of explicit and implicit relations. {The relation is derived from components across artifacts, toolchains, applications, platforms, and other related elements.}


\subsubsection{Explicit Relations}

Explicit relations are those that can be directly identified from artifacts such as source code, configuration files (\eg package manager configuration files), and documentation.

\textbf{Invocation Relations from Source Code.} Explicit dependencies are often revealed through import statements or direct method calls. These include class or method imports, method invocations, the use of imported constants, and API invocations.

\textbf{Dependency Declaration in Config Files.} Dependencies explicitly declared in configuration files, such as \texttt{package.json} in JavaScript or \texttt{pom.xml} in Java, provide clear indications of relationships between components. These files serve as a formal record of required libraries or modules.

\textbf{Dependency Statements in Documentation.} Explicit relations can also be derived from documentation, such as API references, README files, or inline comments, which explicitly describe dependencies, interactions, or usage guidelines. 

\subsubsection{Implicit Relations}

Implicit relations are those embedded during the development and deployment of LLMs. Unlike explicit relations, they cannot be directly inferred from existing artifacts. Nevertheless, they are an inevitable and integral component of the LLM supply chain.

\textbf{Training Relations from Training and Fine-tuning Dataset.} Data stands as one of the most critical components in training LLMs. The direct dependencies on data are consolidated into the trained weights, which are inherently embedded within the models. While explicit patterns are directly extracted, implicit relations, though not directly extractable, can still be inferred through techniques such as membership inference attacks~\cite{feng2024exposing, duan2024membership}.

\textbf{Prompt Augmentation Relations from RAG Dataset and Third-party Prompts.} The RAG dataset serves as an external knowledge source, enabling LLMs to acquire a more comprehensive understanding of the queried questions. Meanwhile, third-party prompts offer detailed and tailored instructions, making it easier for LLMs to interpret and respond effectively. These elements act as supplementary knowledge sources, providing richer contextual information for LLMs. As a result, we define these relationships as prompt augmentation relations.

{\textbf{Artifact Reusing Relations.}} In traditional software development, code reuse commonly occurs through a well-known practice called code cloning. As we transition to the LLM supply chain and encounter new types of artifacts, novel forms of reuse have emerged. For example, a most notably new type is model reuse. To fulfill specific requirements, downstream users may split, merge, or convert models, reflecting more dynamic and complex reuse patterns. Moreover, datasets often undergo preprocessing as a standard part of the pipeline.

\begin{tcolorbox}[size=title, opacityfill=0.15]
    \textbf{Summary.} We explore both explicit and implicit relations within the LLM supply chain. Implicit relations, in particular, deserve significant attention due to their inherent complexity in inference and their pervasive presence across the LLM supply chain. These represent compositional relationships that may introduce new risks and challenges.
\end{tcolorbox}

\section{Risks of Large Language Model Supply Chain (RQ2)}\label{attacks}

We introduce the notion of \textit{risky scenarios}. Then, we illustrate risk types, risky stakeholders, risky actions, respectively.

\subsection{Risky Scenarios}

\begin{figure*}[!t]
    \centering
    \includegraphics[width=0.95\textwidth]{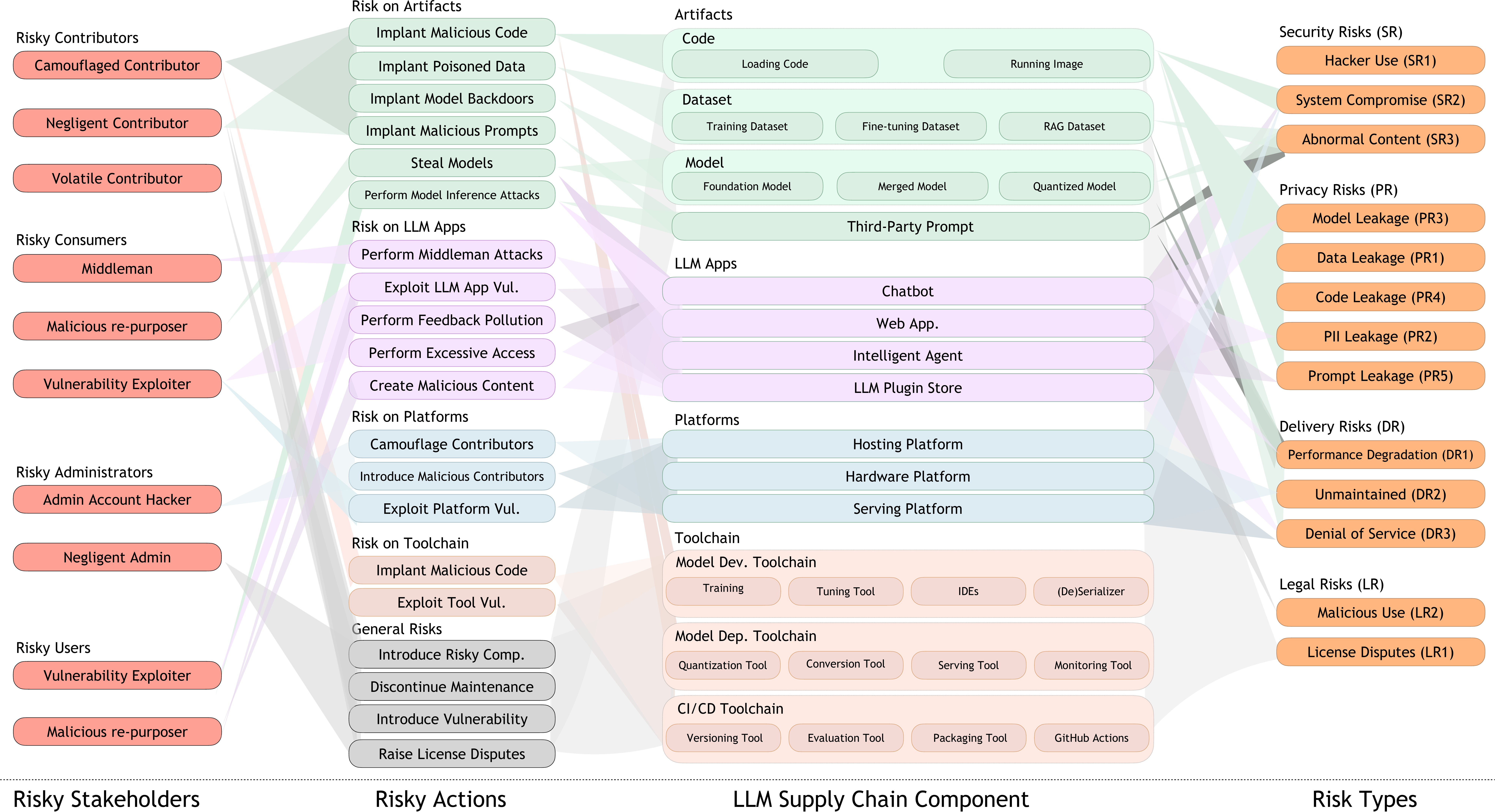}
    \caption{Risky Actions, Supply Chain Components and Risk Types Introduced by Risky Stakeholders}
    \label{fig:riskoverview}
\end{figure*}

The risky scenarios can be illustrated that a group of \textit{risky stakeholders} perform some \textit{risky actions} on specific \textit{supply chain components}, causing certain \textit{risk types}. The \textit{risky stakeholders}, \textit{risky actions}, \textit{supply chain components}, and \textit{risk types} are presented from the first column to the fourth column in Figure \ref{fig:riskoverview}, respectively.
    
\subsection{Risk Types}\label{sec:risk_type}
    
\begin{figure}[!t]
    \centering
    \includegraphics[width=0.80\textwidth]{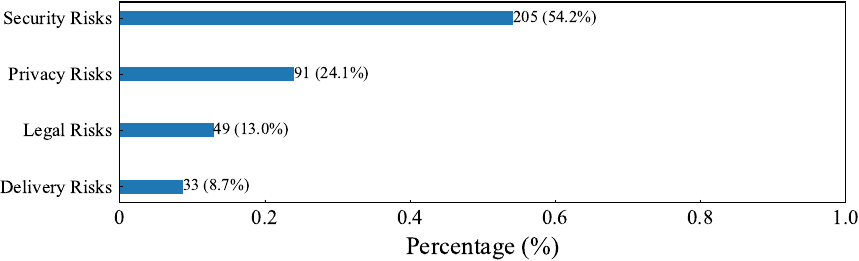}
    \caption{Distribution of Risk Types that Covered in our Collected Literature}
    \label{fig:risktypebar}
  \end{figure}

Figure \ref{fig:risktypebar} illustrates the distribution of risk types addressed in the surveyed literature. A majority of the studies (205 papers, 54.2\%) concentrate on security risks, followed by privacy risks (91 papers, 24.1\%). In contrast, legal risks (49 papers, 13.0\%) and delivery risks (33 papers, 8.7\%) have received comparatively less attention. This uneven distribution calls for greater attention to legal and delivery risks, which remain underexplored despite their potential impact on the LLM supply chain.

  \begin{figure}[!t]
  \centering
  \includegraphics[width=0.80\textwidth]{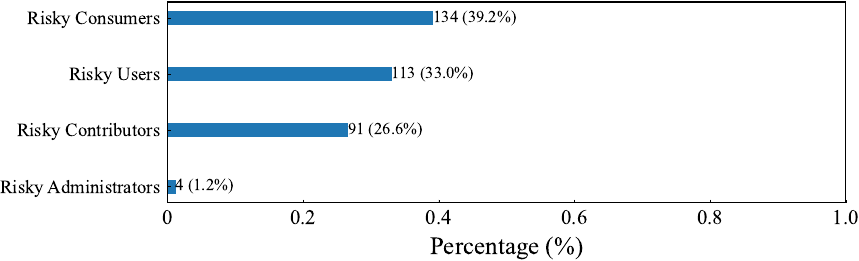}
  \caption{Distribution of Stakeholders that Covered in our Collected Literature}
  \label{fig:stakeholder}
\end{figure}

\begin{itemize}[leftmargin=*]
    \item \textbf{Security Risks \wss{(SR)}} include potential threats such as \wss{hacker use} (SR1), \wss{system compromise (SR2), and the emergence of abnormal content (SR3), \ie manipulated content, frad, misinformation, etc} that could disrupt systems or compromise integrity.
    \item \textbf{Privacy Risks \wss{(PR)}} focus on data vulnerabilities, such as data leakage (PR1), personally identifiable information (PII) \wss{leakage} (PR2), model stealing \wss{(PR3)}, code leakage (PR4), and prompt leakage (PR5), all of which can result in the unauthorized access or misuse of sensitive information.
    \item \textbf{Delivery Risks \wss{(DR)}} highlight issues related to the performance and maintenance of services, including performance degradation (DR1), \wss{system unmaintenance (DR2)}, and denial of service (DoS) (DR3) attacks that hinder service availability.
    \item \textbf{Legal Risks \wss{(LR)}} emphasize complications arising from copyright disputes, license disputes \wss{(LR1)}, and malicious use \wss{(LR2)} of systems, which can lead to legal and compliance challenges.
\end{itemize}

\subsection{Risky Stakeholders}


Figure \ref{fig:stakeholder} illustrates the distribution of risky stakeholders addressed that are discussed in the surveyed literature. A majority of the studies (134 papers, 39.2\%) concentrate on risky consumers, followed by risky users (113 papers, 33.0\%), and risky contributors (91 papers, 26.6\%). Few papers discuss the risk introduced by the risky administrators (4 papers, 1.2\%). The results present a potential research gap in understanding and mitigating risks introduced by those who develop, deploy, or manage LLM applications.

\begin{itemize}[leftmargin=*]
    \item \textit{Risky Contributors:} (a) Malicious contributors can either camouflage legitimate contributors to gain system access or penetrate into systems by leveraging system vulnerabilities. (b) Benign contributors may unintentionally introduce vulnerabilities, either through poor coding practices, lack of maintenance, etc. (c) Volatile contributors may suspend their support under certain circumstances.
    \item \textit{Risky Consumers:} Attackers can play the role of artifact consumers who can (a) act as middlemen that can inject malicious content during the exchange, or (b) re-develop existing artifact for malicious use. Besides, attack can also (c) be platform consumers who can exploit platform vulnerabilities.
    \item \textit{Risky Administrators:} (a) Attackers can hack into an administrator's account, causing severe consequences given the elevated privileges and control. (b) Benign administrators can inadvertently introduce risks into the system through misconfigurations or other unintentional actions.
    \item \textit{Risky Users:} Users whose purposes are to exploit LLM application vulnerabilities, steal models, conduct jailbreaking or inference attacks, perform excessive access, or pollute the data using user feedbacks.
\end{itemize}

\subsection{Taxonomy of Risky Actions}\label{sec:action}

\begin{figure}[!t]
  \centering
  \includegraphics[width=0.80\textwidth]{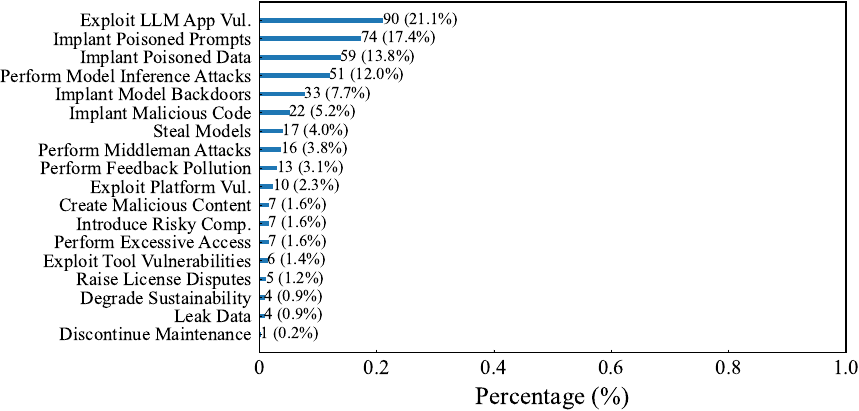}
  \caption{Distribution of Risky Actions that Covered in our Collected Literature}
  \label{fig:riskactions}
\end{figure}

Figure~\ref{fig:riskactions} illustrates the distribution of risky actions addressed in the surveyed literature. The five most frequently discussed actions are exploiting LLM application vulnerabilities (90 papers, 21.1\%), implanting poisoned prompts (74 papers, 17.4\%), implanting poisoned data (59 papers, 13.8\%), performing model inference attacks (51 papers, 12.0\%), and implanting model backdoors (33 papers, 7.7\%). These actions are largely inherent to the unique characteristics of LLMs and represent threats that are distinct from those observed in traditional software supply chains. In contrast, only a few papers address actions such as degrading sustainability (4 papers, 0.9\%), data leakage (4 papers, 0.9\%), and discontinuation of maintenance (1 paper, 0.2\%). Given the continuous operation and evolution of LLM applications, this limited attention to long-term risks may pose significant challenges in the long run.

\subsubsection{Risks on Artifact} The artifact includes code, data, model and prompt.

\textbf{Implant Malicious Code.} Attackers can develop malicious code and attempt to implant it through various methods. For instance, Zhao et al. \cite{zhao2024models} revealed 9 malicious dataset loading scripts on Hugging Face. \wss{It can cause the risk of malicious use (LR2)}.

\begin{itemize}[leftmargin = *]
    \item \textit{Dependency Confusion.} \huangyiheng{Attackers publish a malicious package to a public package repository with the same name as a popular private package. The package manager may unknowingly download the malicious package from the public repository instead of the intended private repository \cite{pytorch_confusion}.}
    \item \textit{Typosquatting.} \huangyiheng{Attackers upload malicious packages with names containing minor typos to imitate popular legitimate packages. Accidentally installing those packages can lead to security issues and compromise the integrity and security of systems \cite{typosquatting}.} 
    \item \textit{Compromising Legitimate Libraries.} \huangyiheng{Attackers may target legitimate libraries already integrated into the LLM supply chain. They can conceal their malicious code within pull requests. In addition, attackers may take control of the code from a previous maintainer and release new malicious versions \cite{event-stream}.}
    \item \textit{Sharing Cloud Environments.} \huangyiheng{Attackers can inject malicious code into cloud environments and make it publicly available to all cloud users. It was discovered that over a thousand Docker container images were found hiding malicious content \cite{docker_malicious_images}. Tomar et al.~\cite{tomar2020docker} proposed a taxonomy of potential attacks at the container layer.}
\end{itemize}

\textbf{Implant Poisoned Data.} Attackers can create poisoned data and implant it during critical stages of building the LLM supply chain. \wss{It can cause the risk of data leakage (PR1) and malicious use (LR2)}.

\begin{itemize}[leftmargin = *]
\item \textit{Training Data Poisoning.} Attackers can manipulate the data for malicious purposes, and then upload it to public platforms (\eg Kaggle)~\cite{ma2019explaining}. Once integrated into the LLM supply chain, it would result in the model performance degradation~\cite{carlini2024poisoning, shu2023exploitability,wan2022you,shan2024nightshade, wan2023poisoning}. 
\item \textit{Retrieval Data Poisoning.} Retrieval-augmented generation (RAG) is widely incorporated to provide the LLM with sufficient contextual knowledge. The database can also be poisoned, guiding the LLM to generate specific, attacker-chosen responses~\cite{zou2024poisonedrag, wang2024poisoned}.
\end{itemize}

\textbf{Implant Model Backdoors.} Attackers can manipulate publicly accessible models to alter their behavior for malicious purposes, ultimately affecting downstream users who rely on the compromised models. Generally, the models refer to various components, including the foundation model, vocabularies, tokenizers, embedding layers, and auxiliary models, etc~\cite{liu2024visual,radford2019language,chen2023minigpt}. \wss{It can cause hacker use (SR1), system compromise (SR2), emergence of abnormal content (SR3), privacy risk (PR), data leakage (PR1), personally identifiable information (PII) \wss{leakage} (PR2), model stealing \wss{(PR3)}, code leakage (PR4), and prompt leakage (PR5).}

\begin{itemize}[leftmargin = *]
\item \textit{Component Dependency Attack.} Attackers can manipulate the contents or structures of the composing models of LLMs, affecting the robustness~\cite{shayegani2023jailbreak, cui2024robustness} and causing unexpected consequences\cite{huang2023training}. 
\item \textit{Backdoor attack.} Backdoor attack embeds hidden backdoors in the model during the training process, allowing the compromised model to perform normally on benign samples but altered on the hidden backdoor input~\cite{li2021hidden,chen2021badnl,pan2022hidden}. 

\end{itemize}

\textbf{\wss{Perform Model Inference Attacks.}} Inference attacks aim to exploit LLM models, causing them to inadvertently disclose sensitive information. They include attribute inference attacks and membership inference attacks. Attribute inference attacks involve deducing sensitive information (\eg race, gender, and sexual orientation) from the behavior or responses of LLM models, even if such information is not explicitly included in the training data .~\cite{staabbeyond, pan2020privacy}.  Membership inference attacks aim to predict whether a data sample was included in the training data of LLM~\cite{mattern2023membership}. \wss{It can cause data leakage (PR1), personally identifiable information (PII) \wss{leakage} (PR2), model stealing \wss{(PR3)}, code leakage (PR4), and prompt leakage (PR5).}

\textbf{Implant Poisoned Prompts.} Attackers can conceal harmful or misleading prompts into publicly accessible repositories or datasets, which are  widely shared and reused for downstream users. \wss{It can cause the performance degradation (DR1) of delivery risks}.

\begin{itemize}[leftmargin = *]
\item \textit{Prompt injection.} Prompt injections disguise malicious instructions as benign, exploiting the incapability of LLMs on distinguishing ``good'' and ``bad'' prompts~\cite{greshake2023not, chen2024struq, shayegani2023jailbreak}.

\item \textit{Jailbreaking.} Jailbreaking attempts to bypass safety filters built in the LLMs~\cite{jailbreaking}, making the LLMs responsive to restricted or insecure content~\cite{wei2024jailbroken, deng2024masterkey,zou2023universal,shen2023anything}.

\end{itemize}

\textbf{\wss{Steal Models.}} Adversaries extract sensitive information (\eg model gradients, model parameters) of machine learning models~\cite{gupta2022recovering, zhang2023ethicist,yang2023code,yu2023bag, carlini2024stealing}, which are then used to generate adversarial examples and bypass any~access restrictions~\cite{shen2023anything}. \wss{It can cause privacy risks (PR3)}.

\subsubsection{Risks on LLM Apps}

The LLM app can be a web application, an agent or a chatbot.

\textbf{Perform Middleman Attacks.} A Man-in-the-Middle (MitM) attack involves an attacker intercepting communication between two parties without their knowledge. MitM attack in the LLM application occurs if an attacker gains access to the communication channel, allowing them to alter or steal sensitive information being shared between a user and the AI system~\cite{kang2024exploiting, si2023mondrian}. \wss{It can cause data leakage (PR1), personally identifiable information (PII) \wss{leakage} (PR2), model stealing \wss{(PR3)}, code leakage (PR4), and prompt leakage (PR5).}

\textbf{Exploit LLM Application Vulnerabilities.} The LLM Plugin Store (\eg GPT store~\cite{GPTstore}) enables users to install plugins to enhance large language model functionalities. However, attackers can publish malicious plugins, guiding the model to output incorrect content through prompt injection~\cite{wu2024new}, lead to account takeovers and data leakage~\cite{pluginaccounttakeover, plugindataexposure, plugintrainingdata}. \wss{It can potentially cause the hacker use (SR1), data leakage (PR1) and performance degradation (DR1)}.

\textbf{Perform Excessive Access.} The LLM application may be unresponsive or malfunction due to excessive requests. For example, performing denial of service attacks within the domain of LLMs can significantly increase the energy consumption and latency of the model~\cite{shumailov2021sponge}. \wss{It can cause hacker use (SR1), system compromise (SR2), emergence of abnormal content (SR3), data leakage (PR1), PII leakage (PR2), model stealing \wss{(PR3)}, code leakage (PR4), prompt leakage (PR5),  and performance degradation (DR1)}.

\textbf{Perform Feedback Pollution.} LLM applications typically leverage user feedbacks to improve their performance~\cite{griffith2013policy} (\eg reinforcement learning based on human feedback), which can possibly be disrupted by false feedbacks~\cite{wan2023poisoning,baumgartner2024best}. \wss{It can potentially cause the hacker use (SR1), system compromise (SR2), emergence of abnormal content (SR3), privacy risk (PR)}.

\subsubsection{Risks on Platforms} 

The platforms support various usages for LLM application such as source code management, data hosting, model training, monitoring, and deployment. 

\textbf{Camouflage Contributors.} Attackers can create fake accounts that imitate existing platform accounts (\eg, official accounts) to publish malicious data, code, models, and more. Users may be misled by these repositories, inadvertently using malicious tools to develop or deploy large models. \wss{It can cause the risk of malicious use (SR1)}.

\textbf{Introduce Malicious Contributors.} Attackers can obtain and exploit credentials of existing accounts or API tokens through various methods~\cite{passwd_guessing, pypi_credential_stealing} (\eg social engineering). Once in possession of these credentials, attackers can gain unauthorized access to a range of resources and services associated with LLMs~\cite{freeapi}. \wss{It can cause the risk of malicious use (SR1)}.

\textbf{Exploit Platform Vulnerabilities.} Attackers can identify and take advantage of weaknesses in the underlying platforms. They leverage flaws in cloud services (\eg SAP AI service~\cite{sap_aiflaw}) to exploit cross-session information leakage, or retrieve sensitive tokens found in CI/CD artifacts (\eg GitHub Actions~\cite{github_action_leaking}). \wss{It can cause the risk of hacker use (SR1), data leakage (PR1) and model stealing~(PR3)}.

\subsubsection{Risks on Toolchain} 
Malicious risks can be introduced at both the model development and deployment phases, compromising the integrity and security of the entire pipeline.

\textbf{Implant Malicious Code}. Attackers can compromise the toolchain by injecting malicious code into the toolchain. For instance, "Xcode-Ghost" \cite{XcodeGhost}, a malicious version of Xcode, was distributed through unofficial channels. Developers unknowingly used this compromised version to build their apps, resulting in the inclusion of malicious code in the final applications. It can cause the security risk of system compromise (SR2) and the privacy risk (PR). 

\textbf{Exploit Tool vulnerabilities}. Attackers exploit known vulnerabilities in development tools, frameworks, or CI/CD components that have not been promptly patched. For instance, the vulnerability of Jenkins Core~\cite{jenkinsadvisory}  can lead to remote code execution (RCE) via agent connections. It can cause the security risk of system compromise (SR2), Performance Degradation(DR1) and Denial of Service (DR3) by triggering unexpected errors or system failures.

\subsubsection{Generic Risks to LLM Components}

Apart from risks that are specific to the four types of components, we elaborate on generic risks to LLM components.

\textbf{Introduce Risky Components.} A naive benign contributor may not be fully aware of the hidden vulnerabilities or security flaws in LLM components. When the benign contributor integrates this component, they inadvertently expose the LLM and its users to potential exploitation. \wss{It can cause the risk of hacker use (SR1), data leakage (PR1) and model stealing (PR3)}.

\textbf{Discontinue Maintenance.} When a contributor stops maintaining a component, updates for security patches, bug fixes, and compatibility improvements are halted. The abandoned component contains outdated and incompatible dependencies that are prone to exploitation or reduce model performance and reliability. \wss{It can cause the risk of unmaintained~(DR2)}.

\textbf{Release Vulnerable Code.} The benign contributors may submit code that contains security flaws or exploits without malicious intent, ranging from weak encryption practices, insecure data handling, or exploitable logic errors. \wss{It can cause hacker use (SR1), system compromise (SR2), and the emergence of abnormal content (SR3)}.

\textbf{Raise License Disputes.} \huangyiheng{While open-source licenses grant general permissions, the owners retain the copyright of the artifact. It introduces potential risks for users~who download and use the code, model or data if contributors claim copyrights on their code. They may enforce restrictions or take legal action if users violate the terms of the license~\cite{CoKinetic_Panasonic}. \wss{It can cause the risk of license disputes (LR1)}.}

\begin{tcolorbox}[size=title, opacityfill=0.15]
    \textbf{Summary.} We illustrate the risky scenarios, including high-risk stakeholders, actions, risk types, and their corresponding components within the supply chain. Compared to the traditional software supply chain, the LLM supply chain involves unique risks, such as \textit{implanting poisoned data}, \textit{model stealing}, and \textit{model inference attacks}. Some risk types are exclusive to LLMs, such as \textit{model leakage}. The affected components in the LLM supply chain represent a new range of software elements that deserve further investigation and consolidation.
    
\end{tcolorbox}

\section{Mitigation of Risks in Large Language Model Supply Chain (RQ3)}\label{defense}

\begin{figure*}[!t]
    \centering
    \includegraphics[width=0.55\textwidth]{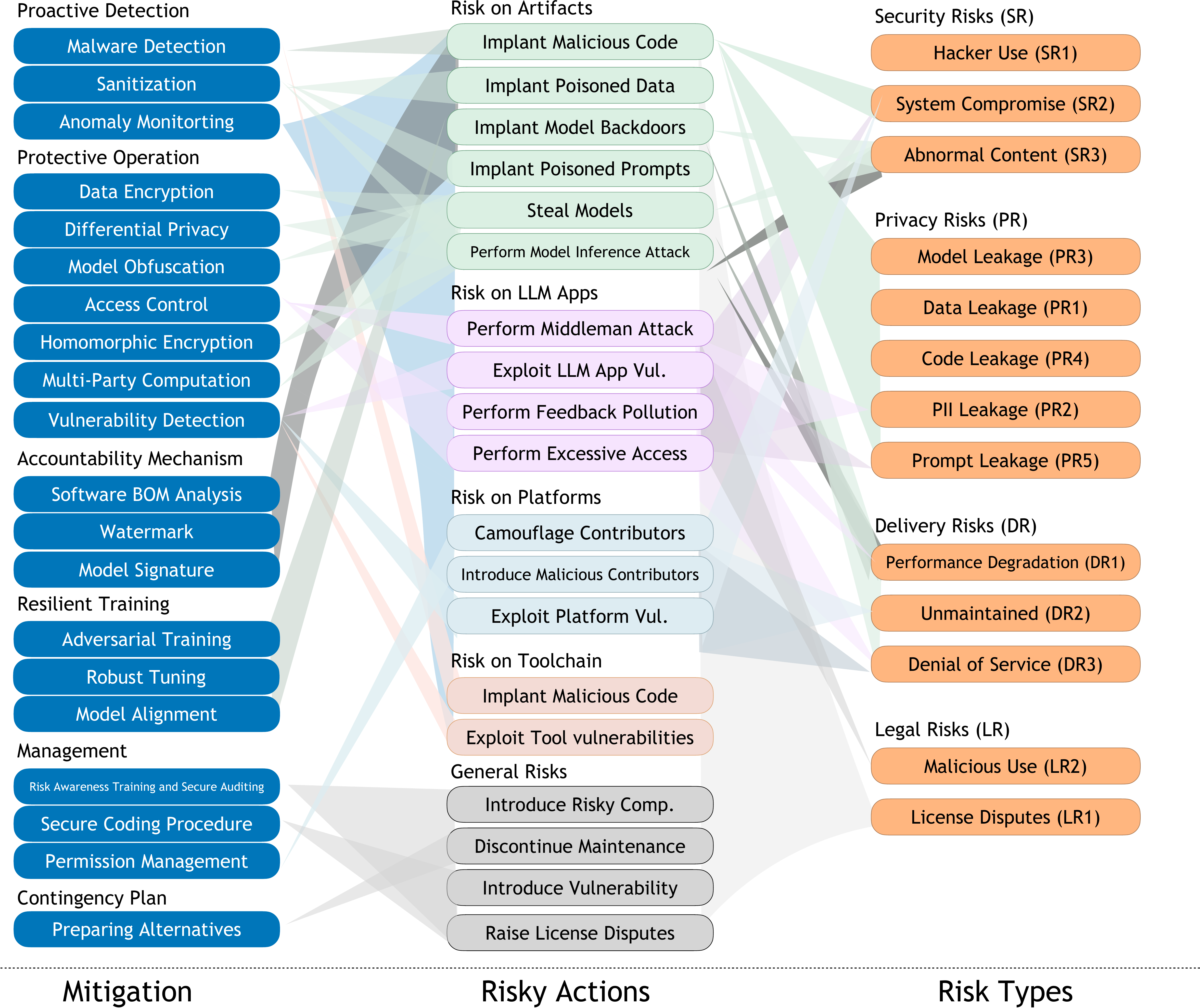}
    \caption{Mitigations against Risky Actions and Risk Types}
    \label{fig:mitigationoverview}
\end{figure*}

We elaborate on the mitigation of risks in the LLM supply chain with regard to six aspects. Figure~\ref{fig:mitigationoverview} presents the distribution of mitigation strategies discussed in the surveyed literature. The majority of studies propose solutions based on proactive detection (74 papers, 33.2\%) and protective operation (56 papers, 25.1\%). Resilient training techniques also account for a significant portion, reflecting efforts to enhance model robustness during development. The fourth and fifth most discussed strategies involve \textit{management} (20 papers, 9.0\%) practices and \textit{accountability mechanisms} (17, 7.6\%), which primarily address socio-organizational aspects of risk mitigation. Notably, only one paper discusses the use of contingency planning as a mitigation approach.

\begin{figure}[!t]
  \centering
  \includegraphics[width=0.80\textwidth]{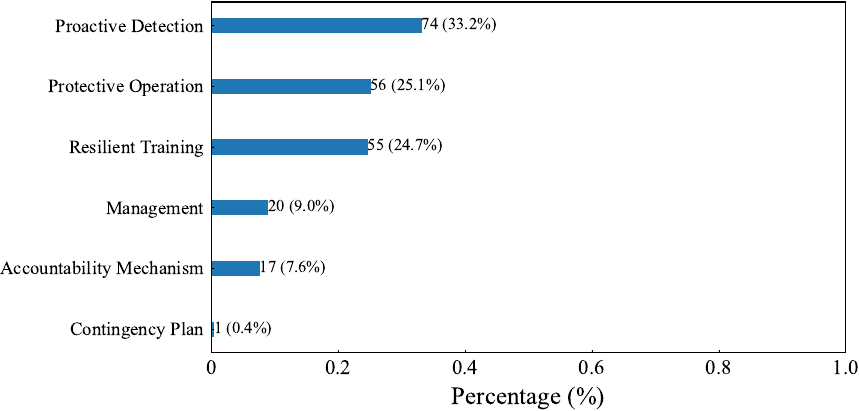}
  \caption{Distribution of Mitigations that Covered in our Collected Literature}
  \label{fig:barmitigation}
\end{figure}

\subsection{Proactive Detection}

\textbf{Malware Detection.} Malware detection identifies viruses, worms, trojans, and software backdoors from third-party libraries in both binaries and source code. Key techniques include signature comparison, static analysis, and dynamic analysis. Signature-based detection tools, like VirusTotal~\cite{virustotal}, are effective for identifying binary malware. Recent research has advanced these approaches by modeling the behavior of malicious packages using static characteristics alone~\cite{SpiderScan, zhang2023cerebro}, as well as integrating both static and dynamic analyses to enhance detection accuracy~\cite{duan2020maloss, huang2024donapi}. This measure can mitigate the risk SR2.

\textbf{Sanitization.} Sanitization involves various techniques to enhance data and prompt security, mitigating risks SR3, PR1, and PR2.

\begin{itemize}[leftmargin=*]
    \item \textit{Data Sanitization.} Data sanitization cleans input data by applying techniques like data denoising, filtering, and smoothing, which effectively remove adversarial noise while retaining legitimate information ~\cite{xu2017feature}. These methods also help prevent data leakage by ensuring that shared or processed data cannot be traced back to individuals. 
    \item \textit{Data Deduplication.} Jagielski et al. discovered that model privacy attacks are more effective on the training dataset that appears multiple times~\cite{jagielski2023measuringforgettingmemorizedtraining}. Removing duplicate text from the training data ~\cite{abadi2024privacy} helps protect against model memorization of sensitive information. 
    \item \textit{Prompt Sanitization.} Prompt sanitization filters, formats, or regularized prompts are provided to LLMs to ensure that the model does not misinterpret their original intent, thereby preventing abnormal content or malicious behavior. 
    Suo et al.~\cite{suo2024signed} applied to sign to sensitive elements (\eg 'delete file') in the user's prompt and Robey et al.~ \cite{robey2023smoothllm} randomly perturb characters in a given input prompt (e.g., inserting random characters), effectively prevent jailbreaking in LLMs. Chen~et~al~.\cite{chen2024struq} separates prompts and user data into two independent channels to prevent prompt injection. 
\end{itemize}

\textbf{Anomaly Monitoring.} Platforms can apply authorized and restricted access for untrusted accounts. \eg multi-factor authorization~\cite{pypi_2fa}. Apart from that, they can adopt anomaly monitoring by observing developing activities to identify any suspicious behavior, such as unauthorized access attempts, changes to account settings, or unexpected pull requests. For example, Danielle et al.~\cite{gonzalez2021anomalicious} leveraged commit messages to detect anomalies and malicious~commits. This measure can mitigate the risk SR2.

\subsection{Protective Operation}

\textbf{Data Encryption.} Data encryption protects data confidentiality by converting it into an unreadable format, consisting of three types: symmetric encryption~\cite{huang2020efficient}, asymmetric encryption~\cite{lin2023memristor,huang2022visually}, and hybrid encryption~\cite{abdelnapi2016hybrid}. With effective key management mechanisms, the data cannot be deciphered, significantly reducing the risk of data privacy~breaches. This measure can mitigate the risk PR1 an~PR2.

\textbf{Differential Privacy.} Differential privacy~\cite{bagdasaryan2020backdoor} protects the private information of individual records in a dataset~\cite{dwork2006calibrating,dwork2014algorithmic}. Specifically, a differential privacy mechanism adds noise to each training data point. It ensures that even if a particular data point is replaced or removed, the model's output will not be significantly affected, which effectively prevents attackers from inferring private information in the training data by analyzing the model's outputs. This measure can mitigate the risk PR2.

\textbf{Access Control.} Access control mechanisms are essential measures to ensure systems secure interactions and protect sensitive information in systems.

\begin{itemize}[leftmargin=*]
    \item \textit{Account Access Control.} Authorization (\ie restricting access to authorized users), authentication (\ie enforcing mutual authentication and encryption), and rate Limiting (\ie limiting access to verified users via rate-limiting mechanisms) ensure authorized clients, credible feedback and controlled access rate to prevent system abuse (\eg Denial of Service (DoS) attacks). This measure can mitigate the risks DR1, DR2 and DR3.
    \item \textit{Session and Data Isolation.} Session isolation techniques (\eg sandboxing) separates user sessions, preventing cross-session attacks. Besides, platforms can utilize secrets management tools, log scanning to ensure that sensitive information like tokens and API keys are not exposed in public repositories or logs~\cite{github_action_leaking}. This measure can mitigate the risks SR2, PR1, PR2, PR3 and PR4.
\end{itemize}

\textbf{Model Obfuscation.} Model obfuscation protects the security and privacy of the LLMs. It modifies the model's structure or parameters to prevent attackers from being reverse-engineered or maliciously exploited. Zhou et al.~\cite{zhou2023nnsplitter} proposed an active defense solution for model theft via automated weight~obfuscation. This measure can mitigate the risk PR3.

\textbf{Homomorphic Encryption.} Homomorphic encryption is a form of encryption that allows computations to be performed directly on encrypted data without needing to decrypt it first. Chen~et~al~\cite{chen2022x} employs homomorphic encryption to enable privacy-preserving inference for Transformer-based models. This measure can mitigate the risk PR1.

\textbf{Multi-Party Computation.} Li~et~al.~\cite{li2022mpcformer} achieve private transformer inference using Secure Multi-Party Computation. Federated learning can be regarded as a particular type of multi-party computation. It enables multiple clients collaboratively train a shared model without transferring their individual data to a central server~\cite{li2020learning,zhang2022fldetector} while keeping the shared model remain secure and resistant to backdoor attacks or malicious data manipulation. This measure can mitigate the risk PR1, PR3 and SR3.

\textbf{Vulnerability Detection.} Vulnerability detection identifies weaknesses or flaws in software or systems that could be exploited by attackers. For example, platforms should also ensure that their systems are following best security practices, such as least privilege, secure credential storage, and network segmentation, keeping all CI/CD tools and dependencies up to date with the latest security patches to protect against known vulnerabilities~\cite{sap_aiflaw}. This measure can mitigate the risk SR2.

\subsection{Accountability Mechanism}

\textbf{Software Bill-of-Materials Analysis.} A Software Bill of Materials (SBOM) is a list of inventory components included in imported third-party libraries, helping users avoid unvetted or potentially risky libraries. Several commercial tools offer SBOM analysis, including Sonatype Management \cite{Sonatype_Managemnent}, BlackDuck \cite{BlackDuck}, and Microsoft's SBOM tool \cite{Microsoft_sbom}. Given the widespread use of large language model applications, SBOM analysis can be expanded to encompass not only third-party libraries, but also models, data, prompts, and other related elements. This measure can mitigate the risk SR2 and SR3.

\textbf{Watermark.} Watermark techniques embed identifiable information into data or models to ensure traceability, authenticity, and protection against misuse. 

\begin{itemize}[leftmargin=*]
    \item \textit{Data Watermark.} Data watermark protects data privacy by embedding specific watermark identifiers~\cite{sun2023codemark,tang2023did} into the training dataset, thus identifying and tracing the data source. This measure can mitigate the risk PR1.   
    \item \textit{Model Watermark.} Model watermark embeds traceable information into the text generated by LLMs to verify the model's source and authenticity without degrading output quality, such as logits modification~\cite{kirchenbauer2023watermark}, entropy-based embedding~\cite{lee2023wrote}, multi-bit payload~\cite{wang2023towards}, and token sampling~\cite{kuditipudi2023robust}. This measure can mitigate the risk PR3.
\end{itemize}

\textbf{Model Signature.} Model signature helps to track and manage the origin of models. Jiang et al.~\cite{jiang2023empirical} described useful attributes for representing the model signature, such as provenance, reproducibility, and portability. Recording and verifying key attributes, provides a reliable reference point, allowing researchers or engineers to trace and verify the model's training process, thereby determining whether the model has been maliciously modified or compromised. This measure can mitigate the risk PR3.

\subsection{Resilient Training}

\textbf{Model Alignment.} Model alignment ensures that a model's outputs and behaviors align with human values and expectations. Reinforcement Learning from Human Feedback~\cite{ouyang2022training} incorporates human feedback into the model's training process, guiding the model to learn which behaviors are appropriate or~not. This measure can mitigate the risk SR3.

\textbf{Adversarial Training.} Adversarial training is a machine learning technique designed to enhance model robustness by exposing the model to adversarial examples~\cite{xhonneux2024efficient} during training. By learning to recognize and handle these malicious inputs, the model becomes more resilient to attacks and unpredictable behavior. This measure can mitigate the risk SR3.

\textbf{Robust Tuning.} Robust tuning enhances the defense of LLMs against prompt risks by incorporating specific techniques during tuning. For example, Yue~et~al.~\cite{yue2022synthetic} fine-tuning a generative language model with differential privacy to generate privacy-preserving synthetic text. Ozdayi et al.~\cite{ozdayi2023controlling} leverage prompt-tuning to control the extraction rates of memorized content, avoiding attack modifying LLM weights. This measure can mitigate the risk~SR3.

\subsection{Management}

\textbf{Risk Awareness Training and Secure Auditing.} Administrators should be vigilant when components are incorporated into their software systems. On the one hand, they should actively monitor for known vulnerabilities in third-party libraries and frameworks, thus reducing their exposure to potential threats. On the other hand, they should familiarize themselves with licensing agreements, intellectual property rights, and the potential liabilities tied to both open-source and proprietary components. Furthermore, they should conduct regular security audits to assess LLM supply chain component security and compliance. This measure can mitigate the risks SR2 and LR1.

\textbf{Secure Coding Procedure.} Adhering to secure coding practices is fundamental to minimizing software vulnerabilities. This includes following best practices for input validation, implementing robust error handling, and adhering to established secure coding guidelines. By prioritizing these practices, developers can substantially lower the likelihood of introducing security flaws into their applications, thereby safeguarding user data and assets. This measure can mitigate the risk SR2.

\textbf{Permission Management.} Permission management includes enforcing strict vetting processes to ensure that only trusted components are introduced into the system. Additionally, employing role-based access control (RBAC) limits permissions, ensuring that users and LLM supply chain components have only the necessary access to perform their functions, thereby reducing the attack surface. This measure can mitigate the risks SR1, SR3 and LR2.

\subsection{Contingency Plan}

\textbf{Preparing Alternatives.} To mitigate the risks associated with reliance on components that may become unsupported or discontinued, it is important to seek alternative solutions. Moreover, advanced replacement techniques are also needed to help migrate the original component into a new one with the least effort and cost. By diversifying their dependencies and preparing for potential discontinuations, the stability and security of the LLM-driven application can be enhanced. This measure can mitigate the risks SR2, DR2 and DR3.

\begin{tcolorbox}[size=title, opacityfill=0.15]
    \textbf{Summary.} We discusses various strategies to mitigate risks in the LLM supply chain. Key measures include malware detection and Software Bill-of-Materials (SBOM) analysis to address risks from third-party libraries, and privacy-preserving techniques such as data deduplication, encryption, and differential privacy to secure datasets. These strategies address a comprehensive range of risks, ensuring a robust framework for LLM development, deployment, and operation.
\end{tcolorbox}

\section{Case Study}\label{case}



    \textbf{Probllama.} Ollama is an open-source and widely-used framework that allows users to operate large language models. Recently, researchers uncovered 6 vulnerabilities in Ollama \cite{ollama}. The vulnerabilities could allow an attacker to carry out a wide-range of malicious actions, including Denial of Service (DoS) attacks, model poisoning, model stealing, etc. Researchers highlighted that the inherent lack of authentication support in these tools makes them vulnerable when exposed to external environments. The \textit{risk scenario} is that \textit{Model Leakage (PR3)}, \textit{Denial of Service (DR3)} are caused by \textit{Vulnerability Exploiter} \textit{stealing models} and \textit{performing excessive access} on AI models, which includes \textit{Chatbots}, \textit{Web Apps} and \textit{Intelligent Agents}, etc. The corresponding mitigation measures include \textit{Access Control} and \textit{Permission Management} \cite{ollama}.

    \textbf{Ray.} Ray is a distributed computing framework designed to simplify the process of scaling AI applications. It was reported that attackers have been exploiting a missing authentication vulnerability in the Ray AI framework to compromise hundreds of clusters \cite{shadowRay}. By default, Ray does not enforce authentication and does not support any type of authorization methods. Reporters say that hundreds of Ray clusters were hacked via this bug \cite{raysecurityweek}, with the attackers stealing AI models and data, database credentials, password hashes, SSH keys, and OpenAI, HuggingFace, and Stripe tokens. The \textit{risk scenario} is that \textit{Data Leakage (PR1)}, \textit{Model Leakage (PR3)}, \textit{PII Leakage (PR3)} are caused by \textit{Vulnerability Exploiter} \textit{stealing artifacts} including \textit{data}, \textit{model} and \textit{code}, etc. The corresponding mitigation measures include \textit{Vulnerability Detection} and \textit{Access Control} .

    \textbf{``nullifAI'' and Hugging Face Safetensors.} Pickle is a popular Python module for serializing and deserializing ML models. However, Pickle is considered an unsafe data format that allows Python code to be executed during deserialization, which can lead to arbitrary code execution. The research team came upon two Hugging Face models containing malicious code exploiting the unsafe mechanisms of Pickle, named as ``nullifAI'' \cite{malicioushf}. The \textit{risk scenario} is that \textit{System Compromise (SR2)} and \textit{Privacy Risks (PR)} are caused by \textit{Vulnerability Exploiter} \textit{exploiting artifact vulnerabilities} including \textit{data}, \textit{model} and \textit{code}, etc.
    
    Furthermore, the Hugging Face introduces a new safe serialization format called Safetensors \cite{safetensor} to mitigate the supply chain risk posed by vulnerable serialization formats. They created a conversion service to convert any PyTorch model contained within a repository into a Safetensors alternative via a pull request. However, researchers demonstrate that attackers could compromise the Safetensors conversion space and its associated service bot \cite{safetensor}. The \textit{risk scenario} is that \textit{System Compromise (SR2)} and \textit{Privacy Risks (PR)} are caused by \textit{Vulnerability Exploiter} \textit{exploiting platform vulnerabilities}. The corresponding mitigation measures include \textit{Vulnerability Detection} and \textit{Malware Detection}.

    \textbf{DeepSeek Brand Impersonation.} DeepSeek AI chatbot quickly gained international attention, making it a prime target for abuse. Attackers leverages a tactic known as brand impersonation, creating look-alike websites to deceive users and steal private information \cite{ds1, ds2}. They also uses a fake CAPTCHA page to deceive users into executing a malicious command. The \textit{risk scenario} is that \textit{System Compromise (SR2)} and \textit{Privacy Risks (PR)} are caused by \textit{Vulnerability Exploiter} \textit{exploiting LLM App. vulnerabilities}, and \textit{Negligent User} \textit{introducing Risky Components}. The corresponding mitigation measures include \textit{Risk Awareness Training and Secure Auditing}, \textit{Software BOM Analysis} and \textit{Malware Detection}.
    

\section{Discussion}\label{discuss}

We discuss future challenges and opportunities towards mitigating the risks in the LLM supply chain.

\subsection{Future Challenges}

\textbf{Comprehensive and Precise Decomposition of the LLM Supply Chain.} Decomposing the LLM supply chain is essential in understanding and mitigating risks in the LLM-related ecosystem, ensuring compliance, and enhancing transparency across the LLM lifecycle. First, \textit{mapping an entire LLM supply chain is a complex task}. As the scope and applications of LLMs expand, defining clear boundaries for the LLM supply chain becomes nearly impossible. Nevertheless, identifying a clear scope of relevant components is crucial to enable a comprehensive analysis of the LLM supply chain, including its core artifacts, stakeholders, and supply relationships. This clarity allows each stakeholder to understand their role and responsibilities within the supply chain effectively. Second, \textit{the unique dependencies of the LLM supply chain adds further complexity}. Unlike traditional software supply chains, the LLM supply chain includes special dependencies—such as models embedding external knowledge (e.g., data or code). The models do not directly use external resources. Instead, they encode knowledge into tensors and weights, obscuring the origins and composition of these elements. Third, \textit{LLM artifacts} (\eg models, datasets, and prompts) \textit{differ fundamentally from software artifact, making it difficult to directly apply standard version management}. This limitation complicates efforts to decompose the LLM supply chain effectively and hinders accurate tracking and~management.

\textbf{Compliance with Government Laws and Guidelines.} The growth of new LLM applications is expected to accelerate with the increasing availability of commercial and open-source LLMs, leading to their adoption across a wide range of industries from different countries. Since industries across countries may be subject to varying regulations and requirements, ensuring compliance with government laws and guidelines is essential. First, as LLMs often rely on data sourced from various online platforms, \textit{verifying the legality of training datasets to see if it is intellectual property (IP) free is essential}. Given that the conclusion of dataset composition may be indirect, it is challenging to establish a common sense for copyright infringement or IP violations solely based on the model's generated output or trained weights. Furthermore, the providers may deliberately wipe traces of the IP for their benefit. Therefore, it is also crucial to fight against those actions. Second, \textit{the vast data required to train LLMs includes sensitive information that may pose privacy risks}. Ensuring compliance with data privacy laws like GDPR, alongside securing data through encryption and anonymization is also a critical issue. Third, \textit{LLMs are prone to inheriting biases present in their training data}, which can persist in the downstream application, affecting their performance and trustworthiness. Developing robust strategies for identifying and mitigating biases is crucial for ethical AI.
    
\textbf{Vigilant Prevention of Adversarial Attacks.} On one hand, with reliance on third-party data sources, libraries, proprietary frameworks, and cloud providers, \textit{there is a risk of supply chain vulnerabilities and limited control over model updates}. This creates security concerns as well as challenges in maintaining compatibility and transparency. On the other hand, \textit{LLMs are increasingly targeted by adversarial attacks}, such as data poisoning or model inversion, which can compromise data integrity or privacy. Developing robust defenses will be necessary to ensure secure usage.

\subsection{Opportunities}

\textbf{Collaborative Open-Source Ecosystem.} The prosperity of LLMs is inherently linked to the open-source community. The techniques for effectively leveraging the open-source ecosystem around LLMs are still underdeveloped. Open-source contributions should incorporate collaborative risk assessments and secure coding practices to prevent vulnerabilities. Additionally, establishing ethical guidelines for contributions, such as promoting diversity in training data and addressing bias, can ensure that open-source models are developed responsibly. Furthermore, there is a need for shared benchmarks and evaluation standards that enable developers to compare models transparently, fostering objective performance improvements and driving innovation.

\textbf{Enhanced Transparency and Traceability.}  Improved transparency in sourcing especially in data and models during the whole LLM application developing process can build user trust, ensure accountability in LLM development, and foster ethical AI practices. There are three potential approaches. First, creating detailed audit trails for data usage and transformations in training pipelines can enhance accountability and support error tracking. Second, allowing controlled access to different data and model components supports transparency while protecting sensitive information. Third, tracking changes across model versions, from initial models to refined iterations, enables stakeholders to understand the model's evolution and track back any issues related to specific changes.

\textbf{Regulatory and Standards Development.} Establishing standardized practices and regulatory frameworks for the LLM supply chain can help align practices across different providers, ensuring security, compliance, and interoperability. On the one hand, we~need to define clear guidelines on data handling, including data anonymization and user consent, which can help prevent privacy infringements. On the other hand, regulatory frameworks could include guidelines for reducing bias, ensuring fairness, etc.

\textbf{Improved Defense Mechanisms.} Developing stronger defenses against adversarial threats can enhance the resilience of LLMs against cyber threats and tampering. Three potential approaches can be considered. First, implementing systems to detect unusual activity or output patterns can help identify and mitigate adversarial attacks effectively. Second, utilizing secure delivery channels from an identified contributor can prevent untrusted modifications to LLMs. Third, establishing automated models or library updates can address vulnerabilities in LLM applications promptly, reducing the risk of exploitation.
    

\section{Conclusions}

The rapid growth of large language models has transformed numerous industries, creating an intricate supply chain that may incur potential risks. This paper presents a comprehensive overview of the LLM supply chain. We identify and categorize the risks inherent in this supply chain, framing them through stakeholders, risky actions, risk types, and specific supply chain components. Additionally, we provide a taxonomy of mitigation strategies, offering actionable guidance for stakeholders seeking to navigate and secure the LLM supply chain. We also highlight emerging challenges and opportunities in securing the LLM supply chain, aiming to inspire further research into robust defenses and proactive security measures.

\bibliographystyle{ACM-Reference-Format}
\bibliography{src/reference}

\end{document}